\documentclass[aps,pre,preprint,groupedaddress]{revtex4-1}

\usepackage{amsmath}
\usepackage{amssymb}
\usepackage{graphicx}

\newcommand{\kl}[1]{\left(#1\right)}
\newcommand{\klm}[1]{\left[#1\right]}
\newcommand{\boldP}{\boldsymbol{P}}
\newcommand{\boldb}{\boldsymbol{b}}
\newcommand{\boldbb}{\boldsymbol{bb}}
\newcommand{\boldI}{\boldsymbol{I}}
\newcommand{\boldu}{\boldsymbol{u}}
\newcommand{\boldv}{\boldsymbol{v}}
\newcommand{\boldE}{\boldsymbol{E}}
\newcommand{\boldB}{\boldsymbol{B}}
\newcommand{\boldq}{\boldsymbol{q}}
\newcommand{\boldQ}{\boldsymbol{Q}}
\newcommand{\boldsigma}{\boldsymbol{\sigma}}
\newcommand{\boldPi}{\boldsymbol{\Pi}}
\newcommand{\ppar}{p_{\parallel}}
\newcommand{\pper}{p_{\perp}}
\newcommand{\Tpar}{T_{\parallel}}
\newcommand{\Tper}{T_{\perp}}
\newcommand{\vpar}{v_{\parallel}}
\newcommand{\vper}{v_{\perp}}
\newcommand{\partpart}[2]{\frac{\partial {#1}}{\partial {#2}}}
\newcommand{\partt}[1]{\frac{\partial {#1}}{\partial t}}

\begin{document}

\title{Energy dissipation and entropy in collisionless plasma}

\author{Senbei Du}
\affiliation{Department of Space Science, University of Alabama in Huntsville, Huntsville, AL 35899, USA}

\author{Gary P. Zank}
\affiliation{Department of Space Science, University of Alabama in Huntsville, Huntsville, AL 35899, USA}
\affiliation{Center for Space Plasma and Aeronomic Research (CSPAR), University of Alabama in Huntsville, Huntsville, AL 35805, USA}

\author{Xiaocan Li}
\affiliation{Los Alamos National Laboratory, Los Alamos, NM 87545, USA}

\author{Fan Guo}
\affiliation{Los Alamos National Laboratory, Los Alamos, NM 87545, USA}
\affiliation{New Mexico Consortium, Los Alamos, NM 87544, USA}

\date{\today}

\begin{abstract}
It is well known that collisionless systems are dissipation free from the perspective of particle collision and thus conserve entropy. On the other hand, processes such as magnetic reconnection and turbulence appear to convert large-scale magnetic energy into heat. In this paper, we investigate the energization and heating of collisionless plasma. The dissipation process is discussed in terms of fluid entropy in both isotropic and gyrotropic forms. Evolution equations for the entropy are derived and they reveal mechanisms that lead to changes in fluid entropy. These equations are verified by a collisionless particle-in-cell simulation of multiple reconnecting current sheets. In addition to previous findings regarding the pressure tensor, we emphasize the role of heat flux in the dissipation process.
\end{abstract}

\maketitle

\section{Introduction} \label{sec:intro}

Particle acceleration and heating are common phenomena in plasma. For example, magnetic reconnection converts free magnetic energy into particle energy and thus leads to the acceleration of particles. In this case, the energy source is the free magnetic energy stored in the unstable anti-parallel magnetic field. Another example is turbulence, which results from the nonlinear interaction of structures and waves, and transports energy from large-scale to small-scale flow energy (or in other words, large eddies to small eddies). The energy is eventually dissipated as heat at very small scales, which leads to the heating of plasma. Magnetic reconnection and turbulence are frequently invoked to explain space observations of particle acceleration and the heating of solar corona and solar wind (e.g., Ref \cite{Parker1988, Matthaeus1999, Zank2018}).

It is well known that the typical space plasma has a very long collisional mean free path (e.g., on the order of au, or astronomical unit, in the solar wind near Earth) and thus is considered as collisionless \cite{Marsch1991}. A consequent question is then how to characterize the dissipation process in a collisionless system. An increase in temperature corresponds to plasma heating, but it does not necessarily represent physical dissipation as it might be adiabatic. The pressure-strain interaction was recently suggested as a proxy of dissipation \cite{Yang2017}, but its validity needs to be further tested.

Classic statistical physics tells us that the dissipation process can be described by entropy, which is a nondecreasing function for an isolated macroscopic system according to the second law of thermodynamics. In kinetic theory, the entropy is often related to Boltzmann's H-function \cite{Wolf2009, Guo2017}. The Boltzmann's H-theorem states that the H-function only decreases in the presence of a collision operator. In this sense, entropy is conserved in collisionless plasma. The conservation of entropy has been verified by recent kinetic simulations \cite{Liang2019}.

In fluid dynamics, the entropy is frequently defined as $S \sim \log(p/\rho^{\gamma})$ where $p$ is the pressure, $\rho$ is the density, and $\gamma$ is the adiabatic index (ratio of specific heats). This expression, which we refer to as fluid entropy, has the advantage that it is easy to calculate. However, as we will show in this paper, the fluid entropy may not be a good proxy for entropy in plasma. In Section \ref{sec:entropy}, we show a derivation of fluid entropy as well as its gyrotropic extension from the thermodynamic perspective. From the basic kinetic theory, it is straightforward to derive evolution equations for the fluid entropy, as shown in Section \ref{sec:equation}. These equations suggest several mechanisms that are responsible for the change of fluid entropy. The relation of entropy and previously discussed pressure-strain interaction \cite{Yang2017} is thus established in this study. In addition, we discuss the role of heat flux, which has been neglected in most previous studies. In Section \ref{sec:simulation}, we demonstrate our results by a collisionless particle-in-cell simulation. Finally, a brief dicussion and conclusions are found in Section \ref{sec:discussion}.

\section{Kinetic and fluid entropy}\label{sec:entropy}
\subsection{Kinetic entropy}

Kinetic entropy or Boltzmann entropy is discussed in most textbooks of statistical physics \citep[e.g.,][]{Reif2009}. It is defined through the number of microstates $\Omega$,
\begin{equation}\label{eq:entro-boltz}
  S = k \log \Omega,
\end{equation}
where $k$ is the Boltzmann constant. Equation \eqref{eq:entro-boltz} is the most general and precise definition of entropy. The second law of thermodynamics states that the entropy thus defined is a nondecreasing quantity for an isolated system.

In kinetic theory, another useful quantity is the ``Boltzmann H'' function:
\begin{equation}\label{eq:boltz-H}
  H = \int f \log f d^3x d^3v,
\end{equation}
where $f$ is the particle distribution function. The Boltzmann H-theorem states that the H function of a system is nonincreasing, i.e.,
\[ \frac{d}{dt} H \le 0. \]
Thus the function $-k H$ can be interpreted as the entropy. Indeed, one can show that the decreasing of H function is due to collision \citep[e.g.,][]{Zank2014}. A recent study shows that the Boltzmann entropy or H function, when carefully evaluated in collisionless PIC simulations, is approximately conserved \citep{Liang2019}.

\subsection{Entropy of ideal gas}

In fluid dynamics and MHD, the entropy is frequently defined as the following
\begin{equation}\label{eq:entro-iso}
  S = \frac{3}{2} \log\frac{p}{\rho^{\gamma}},
\end{equation}
where as usual, $p$ is the pressure, $\rho$ is the mass density, and $\gamma$ is the adiabatic index. What is often overlooked is that this definition of entropy stems from the ideal gas equation of state, and may not be valid for magnetized plasmas. For completeness, we show a brief derivation of the expression \eqref{eq:entro-iso}. We start with the thermodynamic relation
\begin{equation}\label{eq:thermo}
  dS = \frac{dQ}{T} = \frac{dE}{T} + \frac{dW}{T},
\end{equation}
where $T$ is the temperature, $dQ$ is the heat, $dE$ is the change in energy, and $dW$ is the work. For ideal gas, we have the equation of state
\begin{equation}\label{eq:eos-ideal}
  pV = NkT,~ \textrm{or}~~ p = nkT.
\end{equation}
where $N$ is the number of particles, and $n$ is the number density. The important property of ideal gas is that the internal energy $E$ is only a function of temperature (but not the volume) with the following relation
\begin{equation}\label{eq:dE-ideal}
  E(T) = \frac{3}{2}NkT = C_v T \quad\Rightarrow\quad \frac{dE}{T} = \frac{C_v dT}{T} = \frac{3}{2}Nk \frac{dT}{T},
\end{equation}
where $C_v$ is the specific heat at a constant volume. Now we use $p$ and $n$ as independent variables, and assuming a fixed number of particles, so that
\[ dT = \frac{1}{nk} dp - \frac{p}{n^2 k} dn;\quad dV = -\frac{N}{n^2} dn. \]
Using the above relations and $dW = pdV$, we find
\[ dS = C_v \frac{dp}{p} - (C_v + Nk) \frac{dn}{n}. \]
Notice that $C_v + Nk = C_p$ the specific heat at a constant pressure, and $\gamma = C_p/C_v$ the adiabatic index, implying that the above relation then becomes
\begin{equation}\label{eq:dS}
  dS = C_v \kl{\frac{dp}{p} - \frac{C_p}{C_v} \frac{dn}{n}} = C_v d\log \frac{p}{n^{\gamma}} = \frac{3}{2} Nk d\log \frac{p}{n^{\gamma}}.
\end{equation}
Thus, Equation \eqref{eq:entro-iso} is recovered except for some constant factors.

\subsection{Entropy of an anisotropic fluid}

One of the key assumptions made in the derivation of the fluid entropy \eqref{eq:entro-iso} is the ideal gas equation of state \eqref{eq:eos-ideal} and \eqref{eq:dE-ideal}. Now we go one step further and consider anisotropic fluid. The simplest plasma model with anisotropic pressure is the CGL model due to Chew, Goldberger \& Low \citep{Chew1956}, where the pressure tensor is assumed to be of the form
\begin{equation}\label{eq:pressure-CGL}
  \boldP = \ppar\boldbb + \pper(\boldI - \boldbb).
\end{equation}
Here, $\boldP$ is the pressure tensor, $\boldI$ represents the identity tensor, and $\boldb$ represents the magnetic field unit vector. The scalar pressure is related to the parallel and perpendicular pressure according to $p = (\ppar + 2\pper) / 3$. When the parallel and perpendicular pressure are the same, $\ppar = \pper = p$, and the pressure tensor reduces to the isotropic pressure $\boldP = p\boldI$. A comprehensive review of the CGL model is found in Ref \cite{Hunana2019}. On neglecting the heat flux, the CGL pressure follows the evolution equations,
\begin{equation}\label{eq:ppar}
  \frac{d\ppar}{dt} + \ppar\nabla\cdot\boldu + 2\ppar\boldbb:\nabla\boldu = 0;
\end{equation}
\begin{equation}\label{eq:pper}
  \frac{d\pper}{dt} + 2\pper\nabla\cdot\boldu - \pper\boldbb:\nabla\boldu = 0.
\end{equation}
See also equations \eqref{eq:pressure-para} and \eqref{eq:pressure-perp} below for more general cases. Here, the $d/dt = \partial/\partial t + \boldu\cdot\nabla$ represents the convective derivative. The classical CGL imposes the same induction equation as in ideal MHD with the motional electric field only,
\[ \partt{\boldB} = \nabla\times(\boldu\times\boldB). \]
Note that $\boldu$ here is the flow velocity, which can be approximated by the ion flow velocity $\boldu_i$. This simplifies the term $\boldbb:\nabla\boldu$ to
\[ \boldbb:\nabla\boldu = \frac{1}{B}\boldb\cdot(\boldB\cdot\nabla\boldu) = \frac{1}{B}\frac{dB}{dt} + \nabla\cdot\boldu. \]
From here, assuming cold electrons, it is straightforward to derive the classical CGL double adiabatic equations
\begin{equation}\label{eq:double-adiabatic}
  \frac{d}{dt} \kl{\frac{\ppar B^2}{\rho^3}} = 0;\quad \frac{d}{dt} \kl{\frac{\pper}{\rho B}} = 0.
\end{equation}
It should be emphasized that the pressure here consists only of ion contributions, and $\rho = \rho_i + \rho_e$ is the total mass density, although the pressure equations \eqref{eq:ppar} and \eqref{eq:pper} are valid for all species (after assuming CGL pressure and neglecting heat flux).

To find the entropy for a CGL fluid, we again use the thermodynamic relation \eqref{eq:thermo}. The equation of state is similar to the ideal gas case, but we need consider separately the parallel and perpendicular pressure/temperature:
\[ \ppar = nk \Tpar;\quad \pper = nk \Tper. \]
The internal energy now relates to both parallel and perpendicular temperature as
\begin{equation}\label{eq:Eparper}
  E = E_{\parallel} + E_{\perp} = \frac{1}{2}Nk \Tpar + Nk \Tper = \frac{3}{2} NkT.
\end{equation}
Although the relation is the same as the ideal gas case when expressed in terms of the isotropic temperature $T = (\Tpar + 2\Tper)/3$, we argue that a single temperature is no longer sufficient to characterize the macrostate of thermodynamic equilibrium---both $\Tpar$ and $\Tper$ are needed. Therefore, we separate the entropy equation into parallel and perpendicular components:
\[ dS_{\parallel} = \frac{dE_{\parallel}}{\Tpar} + \frac{dW_{\parallel}}{\Tpar};\quad dS_{\perp} = \frac{dE_{\perp}}{\Tper} + \frac{dW_{\perp}}{\Tper}. \]
In calculating the parallel and perpendicular work, one should take into account the exchange between parallel and perpencidular energy in addition to the $pdV$ contribution. Using the CGL pressure equations \eqref{eq:ppar} and \eqref{eq:pper} and the induction equation, the work is modified as
\begin{equation}\label{eq:work-parper}
  dW_{\parallel} = \ppar dV + \ppar V\frac{dB}{B};\quad dW_{\perp} = -\pper V\frac{dB}{B},
\end{equation}
so that the combination of the two yields $dW = pdV$ in the limit $\ppar = \pper$. The relation between Equations \eqref{eq:ppar}, \eqref{eq:pper} and \eqref{eq:work-parper} is shown explicitly in the Appendix. Physically, Equation \eqref{eq:work-parper} stems from the conservation of magnetic flux and particles (see Ref \cite{Guo2017} for a more physical derivation). Then, following the same procedure of the ideal gas case, we obtain the parallel and perpendicular entropy as
\begin{equation}\label{eq:dS-parallel}
  dS_{\parallel} = C_{v\parallel} \kl{\frac{d\ppar}{\ppar} - \frac{C_{p\parallel}}{C_{v\parallel}}\frac{dn}{n} + 2\frac{dB}{B}} = C_{v\parallel} d \log \frac{\ppar B^2}{n^{\gamma_{\parallel}}};
\end{equation}
\begin{equation}\label{eq:dS-perp}
  dS_{\perp} = C_{v\perp} \kl{\frac{d\pper}{\pper} - \frac{dn}{n} - \frac{dB}{B}} = C_{v\perp} d \log \frac{\pper}{n B}.
\end{equation}
Here, we have $C_{v\parallel} = Nk/2$; $C_{v\perp} = Nk$, and thus $\gamma_{\parallel} = 3$; $\gamma_{\perp} = 2$ following Equation \eqref{eq:Eparper}. Note that because the $pdV$ term only appears in the parallel work in Equation \eqref{eq:work-parper}, the perpendicular adiabatic index $\gamma_{\perp}$ is not used in the entropy equation. The sum of Equations \eqref{eq:dS-parallel} and \eqref{eq:dS-perp} gives the total entropy:
\begin{equation}\label{eq:dS-total}
  dS = dS_{\parallel} + dS_{\perp} = C_v d\log \frac{\ppar^{1/3}\pper^{2/3}}{n^{5/3}},
\end{equation}
or
\begin{equation}\label{eq:entro-CGL}
  S = C_v \log \frac{\ppar^{1/3}\pper^{2/3}}{n^{5/3}}.
\end{equation}
Thus we recover the early result obtained by Abraham-Shrauner (see Equation (9)--(11) of Ref \cite{Abraham-Shrauner1967}). Under classical CGL assumptions, the parallel, perpendicular, and total entropy are all conserved quantities following the double adiabatic equations \eqref{eq:double-adiabatic}. However, the regular isotropic fluid entropy \eqref{eq:entro-iso} may not be conserved in this scenario using the normal definition $p = (\ppar + 2\pper)/3$. Reference \cite{Abraham-Shrauner1967} also points out that the CGL fluid entropy \eqref{eq:entro-CGL} is consistent with the Boltzmann H-function \eqref{eq:boltz-H} evaluated with a bi-Maxwellian distribution
\[ f = n\kl{\frac{m}{2\pi kT_{\parallel}}}^{1/2}\kl{\frac{m}{2\pi kT_{\perp}}} \exp\klm{-\frac{m\vpar^2}{2kT_{\parallel}} - \frac{m\vper^2}{2kT_{\perp}}}, \]
which can be easily verified.

\section{Plasma energization, heating, and the entropy equations}\label{sec:equation}
\subsection{Plasma energization and heating}

We now use the fluid equations to study the plasma energization and heating. Deriving moments equations from the Vlasov equation is known from most standard plasma textbooks, so it is not shown here (see e.g., Ref \cite{Zank2014, Gurnett2005}). The relevant results are the equations of bulk kinetic energy and thermal energy, as shown in Ref \cite{Yang2017},
\begin{equation}\label{eq:bulk}
  \partt{\varepsilon^{k}} + \nabla\cdot (\varepsilon^{k} \boldu) = -\nabla\cdot(\boldP\cdot \boldu) + \boldP:\nabla\boldu + nq\boldE\cdot \boldu;
\end{equation}
\begin{equation}\label{eq:thermal}
  \partt{\varepsilon^{th}} + \nabla\cdot (\varepsilon^{th} \boldu) = -\boldP:\nabla\boldu - \nabla\cdot \boldq;
\end{equation}
\begin{equation}\label{eq:total}
  \partt{\varepsilon} + \nabla\cdot (\varepsilon \boldu) = -\nabla\cdot(\boldP\cdot \boldu) - \nabla\cdot \boldq + nq\boldE\cdot \boldu.
\end{equation}
Here, $\boldE$ is the electric field and $q$ is the charge of the considered particle species (we drop the species subscript for simplicity). Other quantities in the equations are defined as moments: $n$ and $\boldu$ are the number density and bulk flow velocity as usual; $\varepsilon = (1/2)\int f mv^2 d^3v$ is the total energy density of a plasma species with $f(v)$ the velocity distribution function; $\varepsilon^{k} = (1/2)nmu^2$ is the bulk flow energy density; $\varepsilon^{th} = (1/2)\int fm(\boldv-\boldu)^2 d^3v$ is the thermal energy density. The pressure tensor is defined as the second moment $\boldP = \int fm(\boldv - \boldu)(\boldv - \boldu) d^3v$, and the heat flux vector is $\boldq = (1/2)\int fm|\boldv - \boldu|^2(\boldv - \boldu) d^3v$ (note that it is not to be confused with the charge q).

For an isolated system, one may argue that the divergence terms in Equation \eqref{eq:bulk}--\eqref{eq:total}, when integrated over space, vanish because of Gauss' theorem. Therefore, one arrives at the conclusion that the total plasma energization is due to the work done by the electric field, and the pressure tensor term $\boldP:\nabla\boldu$ acts as a channel that connects the bulk flow energy to the thermal energy \cite{Yang2017}. In other words, it is the work done by the pressure tensor that contributes to the plasma heating.

To further illustrate the physical meaning of the term $\boldP:\nabla\boldu$, the full pressure tensor is decomposed as
\begin{equation}\label{eq:pressure-full}
  \boldP = \ppar\boldbb + \pper(\boldI - \boldbb) + \boldP^{n} = \boldP^{c} + \boldP^{n}.
\end{equation}
Here, we recognize the familiar CGL pressure tensor $\boldP^{c}$ as in Equation \eqref{eq:pressure-CGL} and the remaining part may be called nongyrotropic pressure $\boldP^{n}$ (see Ref \cite{Hunana2019} for a detailed discussion). Upon decomposing the pressure tensor, it is easily verified that
\begin{equation}\label{eq:pressure-work}
  \boldP:\nabla\boldu = p\nabla\cdot \boldu + \frac{1}{2}(\ppar - \pper)\boldbb:\boldsigma + \boldP^{n}:\nabla\boldu,
\end{equation}
where we define the shear tensor $\boldsigma$ as
\begin{equation}\label{eq:shear}
  \sigma_{ij} = \partpart{u_i}{x_j} + \partpart{u_j}{x_i} - \frac{2}{3}\delta_{ij}\nabla\cdot \boldu.
\end{equation}
The shear tensor is equivalent to the traceless strain-rate tensor ($D_{ij}$ of Ref \cite{Yang2017}) within a factor of 2. The first term of Equation \eqref{eq:pressure-work} represents the energization due to fluid compression, the second term represents the shear or viscous energization, and the third term is due to nongyrotropic effects \cite{Li2018, Du2018}. Since the nongyrotropic pressure $P^{n}$ is traceless, the last term can also be written as
\[ \boldP^{n}:\nabla\boldu = \frac{1}{2}\boldP^{n}:\boldsigma. \]
Alternatively, if one considers an isotropic pressure instead, the pressure tensor can be decomposed as
\begin{equation}
  \boldP = p\boldI + \boldPi,
\end{equation}
where $\boldPi$ is the anisotropic pressure tensor (or deviatoric pressure tensor). This decomposition is adopted by e.g., Yang et al. \cite{Yang2017}. Under this decomposition, the pressure tensor work becomes
\begin{equation}
  \boldP:\nabla\boldu = p(\nabla\cdot \boldu) + \frac{1}{2}\boldPi:\boldsigma,
\end{equation}
which is Equation (12) in Ref \cite{Yang2017}. Here, the fluid compression effect is the same as Equation \eqref{eq:pressure-work}, and we find that the pressure-strain interaction (sometimes called ``Pi-D'') term is equivalent to the sum of shear and nongyrotropic energization.

\subsection{The entropy evolution equations}

The discussion above suggests that the plasma heating in an isolated collisionless system is due to the pressure tensor work, which includes compression, shear, and nongyrotropic energization. Another related question is how the heating process relates to actual dissipation and entropy change. In general, an isolated system should conserve entropy when collisions are completely ignored. This is indeed the case for the Boltzmann entropy as illustrated nicely in Ref \cite{Liang2019}. On the other hand, the fluid entropy \eqref{eq:entro-iso} or \eqref{eq:entro-CGL} need not be a conserved quantity. Similar to the derivation of the energy equations, it is straightforward to obtain an entropy evolution equation from moments of the Vlasov equation. For an isotropic fluid, we need use the continuity equation and the equation for scalar pressure. The scalar pressure is related to the thermal energy as $\varepsilon^{th} = (3/2)p$, which is easy to see from their definitions (also note that this is consistent with the ideal gas equation of state \eqref{eq:eos-ideal} and \eqref{eq:dE-ideal}). The result is
\begin{eqnarray}
  \frac{dS}{dt} = \partt{S} + \boldu\cdot\nabla S &=& -\frac{\boldPi:\boldsigma}{2p} - \frac{\nabla\cdot\boldq}{p} \nonumber\\
  &=& -\frac{\ppar - \pper}{2p}\boldbb:\boldsigma - \frac{\boldP^{n}:\boldsigma}{2p} - \frac{\nabla\cdot\boldq}{p}, \label{eq:dSdt-iso}
\end{eqnarray}
where we let
\[ S = \frac{3}{2}\log \frac{p}{n^{5/3}}. \]
Note that we have introduced the convective derivative $d/dt$ on the left side of the equation. Equation \eqref{eq:dSdt-iso} suggests that the entropy change of a fluid element is due to the pressure-strain interaction (or shear/nongyrotropic energization) and the heat flux. Comparing equation \eqref{eq:dSdt-iso} with \eqref{eq:total}, one notices that the compression energization $p\nabla\cdot\boldu$ is absent in the entropy equation, which indicates that adiabatic compression effect does not contribute to the change of entropy. This is because the compression term is absorbed into the convective derivative. However, compression could be important in plasma heating as it is part of the pressure tensor work, and this has been verified by previous simulation studies \cite{Li2018, Du2018}. Another key difference is that there are no divergence terms in the entropy equation. This means that in principle, all terms in Equation \eqref{eq:dSdt-iso} could contribute to the entropy change, even if integrated over the volume of an isolated system.

For cases of anisotropic pressure, we need the equations for both parallel and perpendicular pressure, and they can be derived from the Vlasov equation in a similar way as the scalar pressure equation. Here we take the result from Ref \cite{Hunana2019} (after slight rearrangements),
\begin{equation}\label{eq:pressure-para}
  \partt{\ppar} + \nabla\cdot(\ppar\boldu) + 2\ppar\boldbb:\nabla\boldu + 2\boldbb:(\boldP^{n}\cdot\nabla\boldu) - \frac{d}{dt}(\boldbb):\boldP^{n} + \boldbb:(\nabla\cdot\boldQ) = 0;
\end{equation}
\begin{eqnarray}
  \partt{\pper} + \nabla\cdot(\pper\boldu) + \pper(\boldI - \boldbb):\nabla\boldu + \boldP^{n}:\nabla\boldu - \boldbb:(\boldP^{n}\cdot\nabla\boldu) + \frac{1}{2}\frac{d}{dt}(\boldbb):\boldP^{n} \nonumber\\
  + \nabla\cdot\boldq - \frac{1}{2}\boldbb:(\nabla\cdot\boldQ) = 0. \label{eq:pressure-perp}
\end{eqnarray}
We have introduced the symmetric third rank heat flux tensor $\boldQ = \int fm(\boldv-\boldu)(\boldv-\boldu)(\boldv-\boldu) d^3v$, and it is related to the heat flux vector via a trace operation
\[ \boldq = \frac{1}{2}Tr(\boldQ),\,\, \textrm{or}\,\, q_i = \frac{1}{2}Q_{ijj} = \frac{1}{2}Q_{jij} = \frac{1}{2}Q_{jji}. \]
It can be easily verified that the sum of Equations \eqref{eq:pressure-para} and \eqref{eq:pressure-perp} yields the thermal energy equation \eqref{eq:thermal}. Using the parallel and perpendicular pressure equations, we can derive the evolution equation for the CGL entropy \eqref{eq:entro-CGL},
\begin{equation}
  \frac{dS^c}{dt} = -\frac{\boldP^{n}:\boldsigma}{2\pper} - \frac{\nabla\cdot\boldq}{\pper} - \kl{\frac{1}{\ppar} - \frac{1}{\pper}}\klm{\boldbb:(\boldP^{n}\cdot\nabla\boldu) - \frac{1}{2}\frac{d}{dt}(\boldbb):\boldP^{n} + \frac{1}{2}\boldbb:(\nabla\cdot\boldQ)}, \label{eq:dSdt-CGL}
\end{equation}
where
\[ S^c = \frac{3}{2}\log \frac{\ppar^{1/3}\pper^{2/3}}{n^{5/3}}. \]
On comparing to Equation \eqref{eq:dSdt-iso}, it is clear that the increase of CGL entropy is determined by higher-order moments, namely the nongyrotropic pressure and the full heat flux tensor, while the increase of normal fluid entropy depends also on the gyrotropic pressure and heat flux vector. In the limit of $\ppar = \pper$, \eqref{eq:dSdt-CGL} and \eqref{eq:dSdt-iso} reduce to the same equation. Equation \eqref{eq:dSdt-CGL} is fully general for nonrelativistic collisionless plasmas regardless of the closure.

As is evident from Equations \eqref{eq:dSdt-iso}, for ideal MHD where all third-order or higher moments are cut off and only the isotropic scalar pressure is included, the entropy of a fluid element will be conserved. And similarly for a CGL plasma that satisfies the double adiabatic equations, the CGL fluid entropy is conserved according to Equation \eqref{eq:dSdt-CGL}. We caution that strictly speaking, these conclusions may not be valid in a weak sense. For example, it is well known that shock waves provide dissipation and increase fluid entropy across discontinuous surfaces even within the framework of ideal-gas or ideal MHD (e.g., Ref \cite{Kennel1985}). However, these simple fluid models cannot provide the physical mechanisms that explain the entropy increase and kinetic theory has to be used.

\section{Simulation results}\label{sec:simulation}

In this section, we evaluate the fluid entropy in fully kinetic collisionless particle-in-cell (PIC) simulations, using the code VPIC. It has been recently reported that kinetic Boltzmann entropy is conserved in collisionless PIC simulations \cite{Liang2019}. Fluid entropy \eqref{eq:entro-iso} has also been evaluated in previous PIC simulations such as Ref \cite{Birn2006}, but the detailed mechanisms underlying the increase in fluid entropy remain obscure. We will analyze the change of fluid entropy based on the results presented in the previous section.

We set up a 2D PIC simulation of reconnecting current sheets in a force-free configuration with magnetic field,
\[ B_x = B_0 \tanh\klm{\frac{d}{\pi L}\sin\kl{\frac{\pi z}{d}}} = B_0 \tanh\klm{\frac{\alpha}{\pi}\sin\kl{\frac{\pi z}{\alpha L}}}; \]
\[ B_y = B_0 \sqrt{1 + \kl{\frac{B_g}{B_0}}^2 - \tanh^2\klm{\frac{d}{\pi L}\sin\kl{\frac{\pi z}{d}}}};\quad B_z = 0. \]
Here, $B_0$ is the asymptotic in-plane magnetic field, $B_g$ is the out-of plane guide field, and $L$ is the half thickness of the current sheet. Another parameter $d$ is introduced as the distance between two adjacent current sheets, and $\alpha$ represent the ratio $d / L$. The electric current is calculated according to the MHD force-free condition $(\nabla\times\boldsymbol{B})\times\boldsymbol{B} = 0$. Both electrons and ions follow drifting Maxwellian distributions initially, and we assume the initial density and temperature profiles of both electrons and ions are uniform. The current is carried by the electron drift along the magnetic field. A similar setup has been used in numerous previous simulation studies involving the interaction of multiple current sheets, e.g., Ref \cite{Drake2010, Sironi2011, Hoshino2012}. We set the simulation box $L_x = L_z = 50 d_i$, sheet thickness $L = 0.5 d_i$, guide field $B_g = 0$, and distance $d = 12.5 d_i$ so there are 4 current sheets initially. The simulation is run for $\Omega_{ci}t = 200$, which is about 4 Alfv\'{e}n crossing times $t_{A} = L_x / v_{A}$. 200 macro-particles per cell per species are used in the simulation. A double periodic boundary condition is employed in both the $x$ and $z$ directions, so that the simulation represents a closed system.

\begin{figure*}[!ht]
  \includegraphics[width=1.0\linewidth]{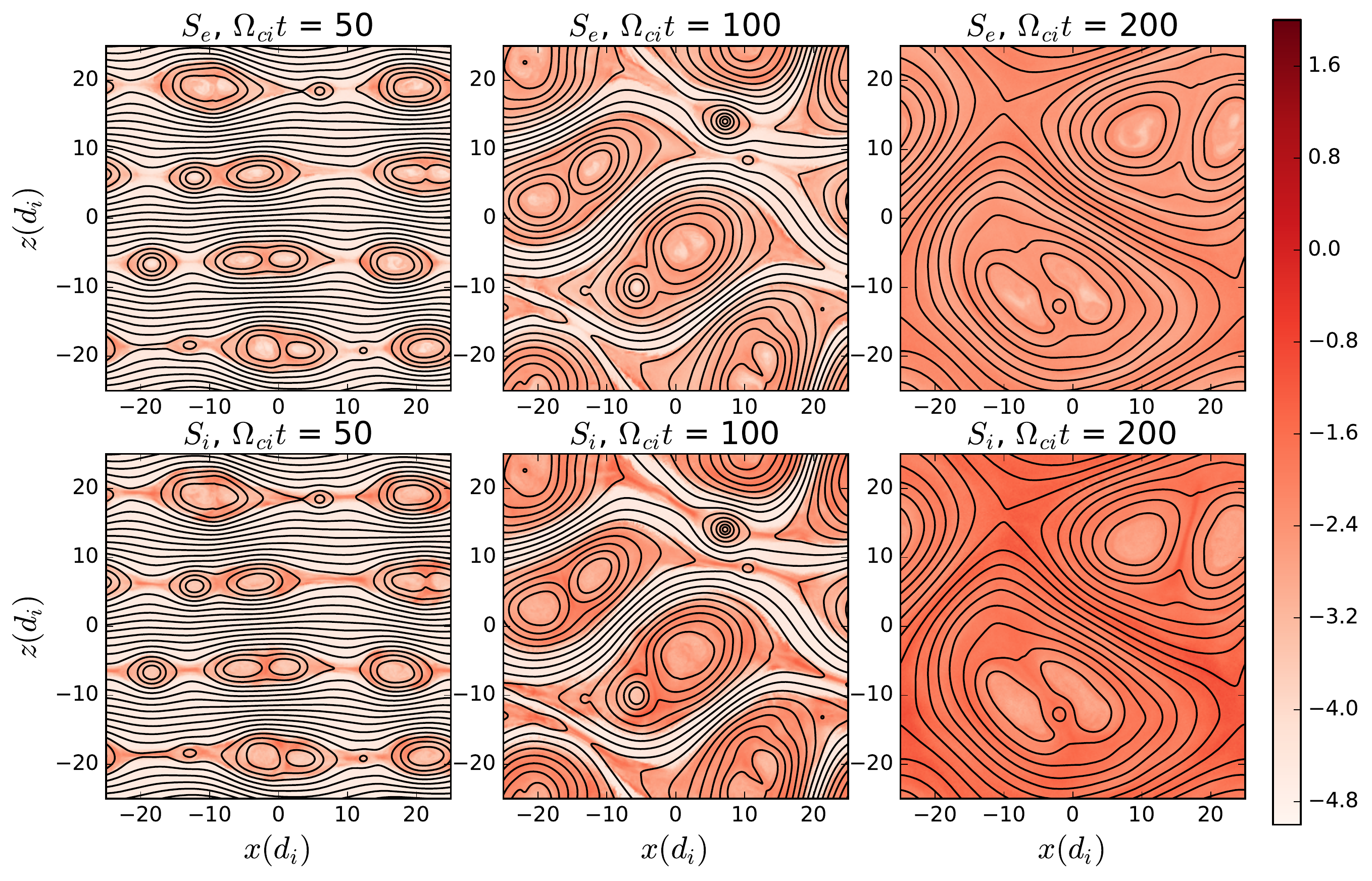}
  \caption{Snapshots of isotropic fluid entropy for electrons (top) and ions (bottom) at different simulation times $\Omega_{ci} t = 50$, 100, and 200. Black contour lines represent magnetic field lines. \label{fig:entro}}
\end{figure*}

Figure \ref{fig:entro} shows snapshots of the isotropic fluid entropy \eqref{eq:entro-iso} in the simulation at different times $\Omega_{ci}t = 50$, 100, and 200. The top panels plot the electron entropy and the bottom panels the ion entropy. Magnetic field lines are overplotted as black contours. Note that the absolute value of the entropy is not important as it depends on how we normalize the pressure and density, but the difference in entropy is independent of normalization. At early stage of the simulation, each current sheet evolves in a relatively independent manner, as illustrated in the left panels ($\Omega_{ci}t = 50$). Numerous small magnetic island structures are formed along each current layer. The current sheets are distorted and interact with each other at later time, illustrated by the middle panels ($\Omega_{ci}t = 100$). The figures suggest that the fluid entropy is produced mostly within the magnetic islands. However, for ions, the entropy seems to be more concentrated near the reconnection X-lines. Approaching the end of the simulation, the magnetic islands have experienced multiple merging and coalescence interactions. The simulation domain becomes rather uniform with a few big island structures.

\begin{figure}[!ht]
  \includegraphics[width=3.375in]{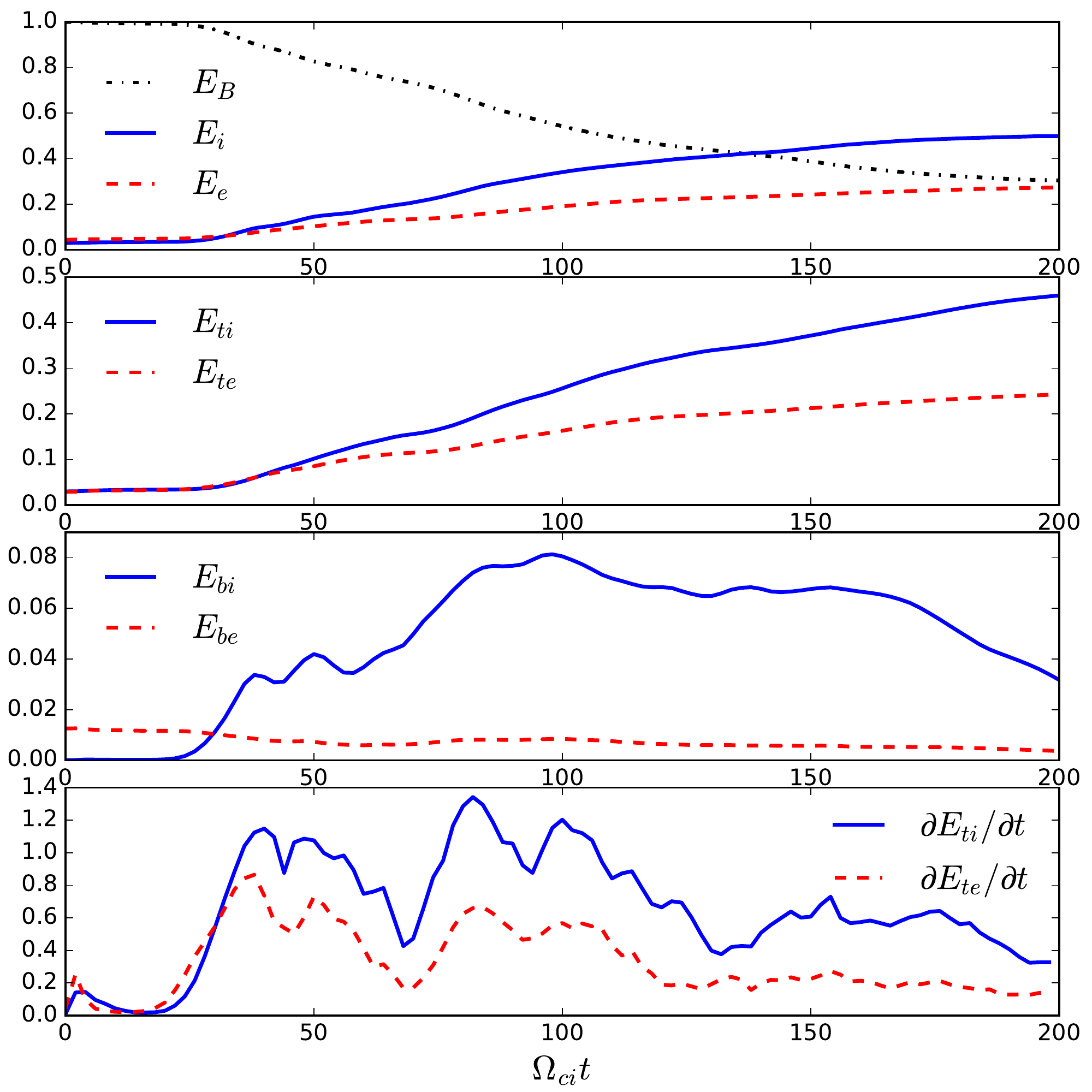}
  \caption{Energy budget of the simulation. The top panel shows the evolution of magnetic energy (dot-dashed black), ion kinetic energy (solid blue), and electron kinetic energy (dashed red). The second panel from top shows the ion (solid blue) and electron (dashed red) thermal energy. The third panel shows the ion (solid blue) and electron (dashed red) bulk flow energy. All energies in these panels are normalized to the initial magnetic energy. In the bottom panel, the rates of change in ion (solid blue) and electron (dashed red) thermal energy are shown. \label{fig:energy}}
\end{figure}

Figure \ref{fig:energy} demonstrates the energy budget of the simulation, normalized to the initial magnetic energy. The top panel shows the evolution of magnetic energy (dot-dashed black curve) and plasma energy (solid blue curve for ions and dashed red curve for electrons). Toward the end of the simulation, about 70\% of the initial magnetic energy is released via magnetic reconnection. Of the released magnetic energy, the majority goes to the ion kinetic energy. The second and third panels show the partition of thermal and bulk flow energy respectively, calculated by
\[ E_t = Tr(\boldP) / 2;\quad E_b = \frac{1}{2}\rho u^2. \]
Again, the solid blue curves represent ions and dashed red curves electrons. The plasma energization is dominated by the thermal energy as more than 90\% the plasma energy resides in the thermal energy at the end of the simulation. The bulk flow energy of electrons is decreasing most of the time during the simulation because the current is carried by electrons initially. The ion bulk flow energy increases at the beginning and is later converted to thermal energy. In the bottom panel, we plot the rate of change in the ion and electron thermal energy.

\begin{figure*}[!ht]
  \includegraphics[width=0.5\linewidth]{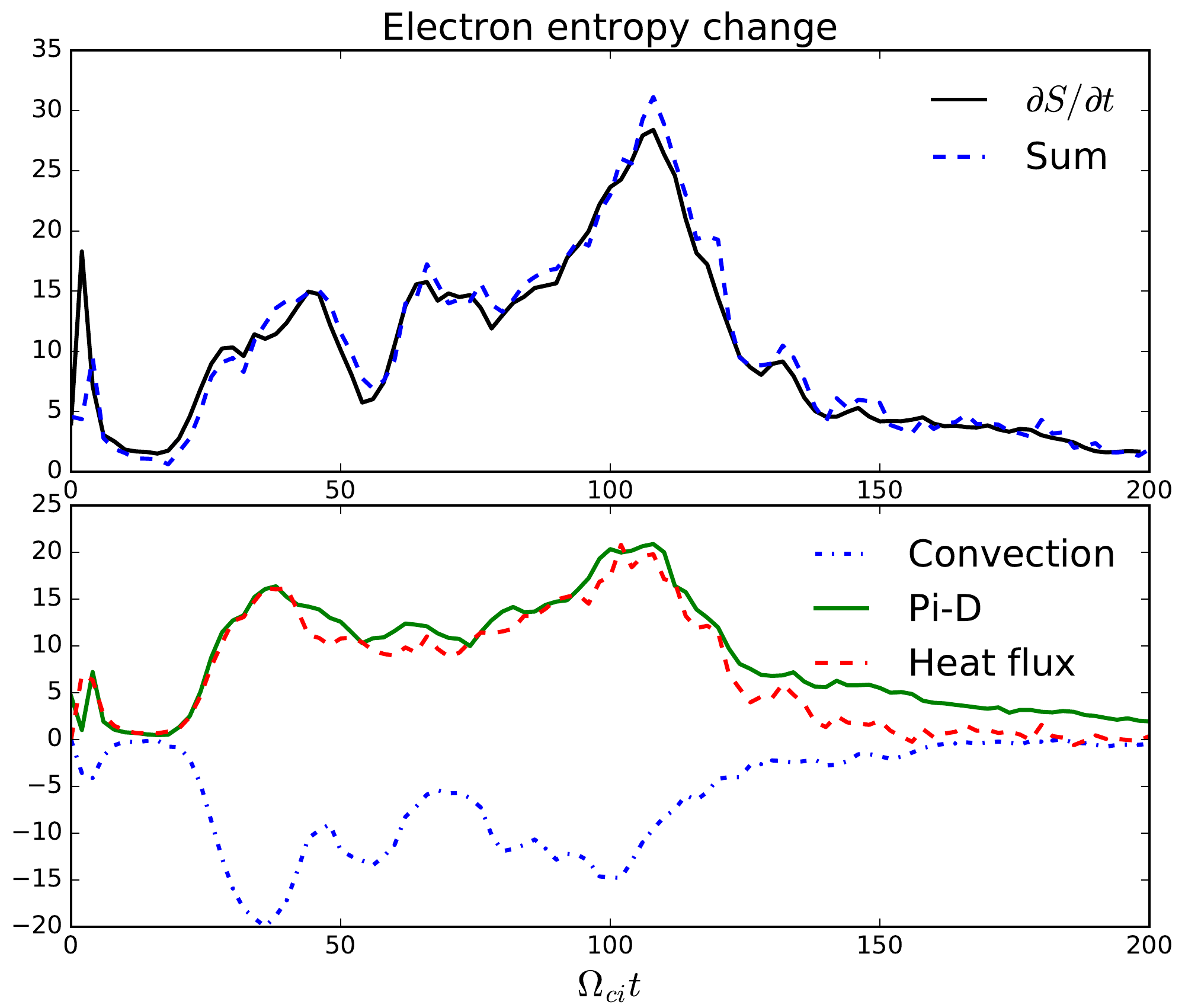}%
  \includegraphics[width=0.5\linewidth]{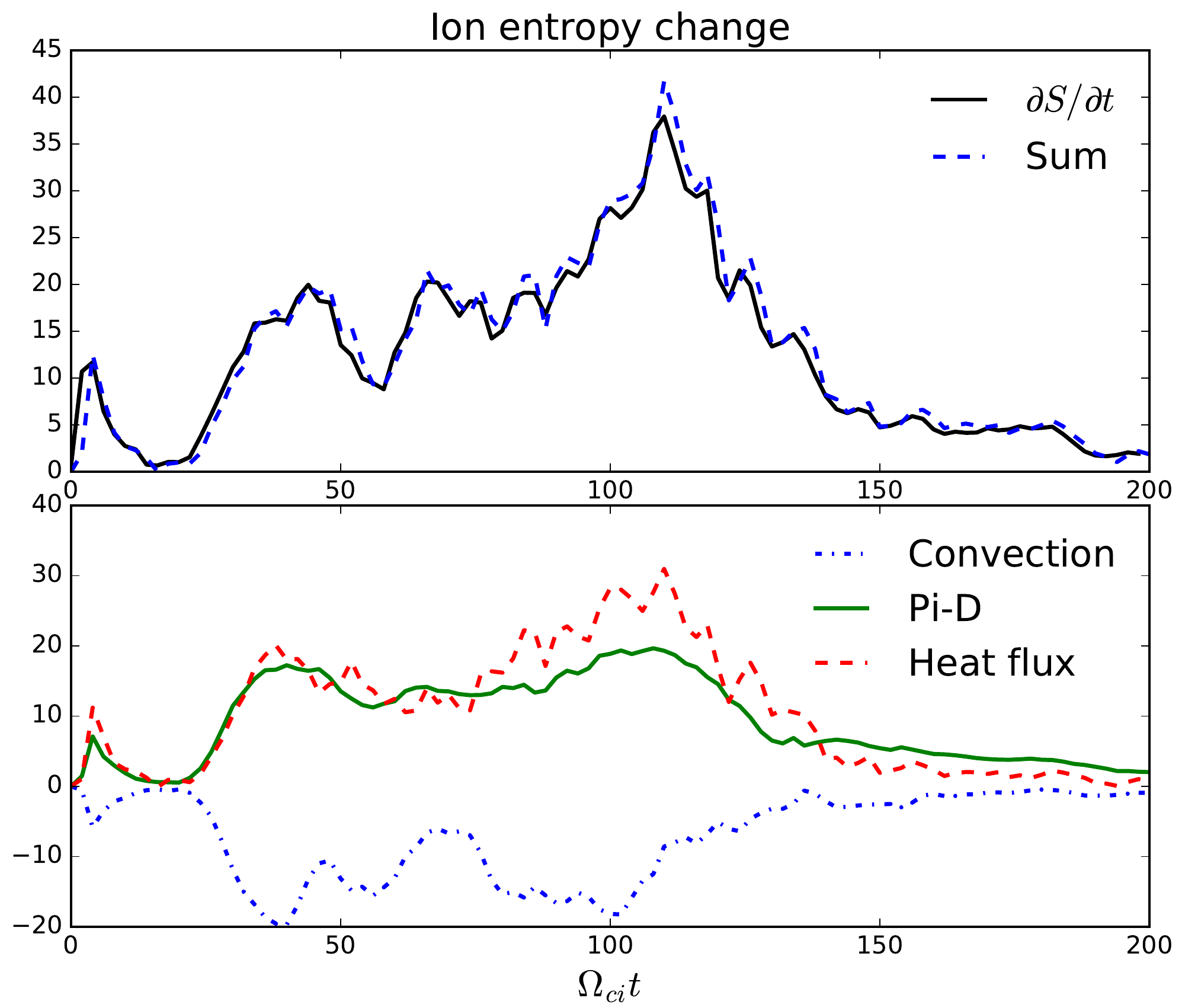}
  \caption{Rate of change of the fluid entropy for electrons (left panels) and ions (right panels). The solid black lines in the top panels represent the rate of change of the fluid entropy. The convection term, Pi-D, and heat flux terms are evaluated separately and shown in the bottom panels. The sum of the three terms is displayed as the dashed blue curves in the top panels. \label{fig:dS-iso}}
\end{figure*}

We then evaluate the rate of change in fluid entropy using Equation \eqref{eq:dSdt-iso} and the result is shown in Figure \ref{fig:dS-iso}. The rate of change of the fluid entropy $\partial S/\partial t$ is integrated over the whole simulation box and plotted in the top panels as solid black lines. The change of fluid entropy is separated into three parts according to Equation \eqref{eq:dSdt-iso}: convection ($-\boldu\cdot\nabla S$), Pi-D ($-\boldPi:\sigma/2p$), and heat flux ($-\nabla\cdot\boldq/p$). They are evaluated for the entire simulation box and are plotted as dot-dashed blue, solid green, and dashed red lines. The sum of the three yields the dashed blue curves in the top panel and it follows the entropy change curve closely, which verifies the Equation \eqref{eq:dSdt-iso}.

It is clear from Figure \ref{fig:dS-iso} that the total fluid entropy of the system is always increasing in our simulation since its rate of change remains positive. This is similar to the thermal energy as illustrated in the bottom panel of Figure \ref{fig:energy} though the two curves do not trace each other exactly. The rate of entropy change is large between $\Omega_{ci}t \sim$ 30--120 when magnetic reconnection and island merging are apparent in the simulation. The overall behaviors of entropy are similar for both electrons and ions. The convection effect remains negative for both species. An interesting result is that the Pi-D and heat flux contributions appear to be comparable with each other. For ions, the heat flux contribution is slightly larger compared with Pi-D. However, at later time of the simulation, the heat flux contribution becomes small and Pi-D dominates the entropy increase for both electrons and ions.

\begin{figure*}[!ht]
  \includegraphics[width=0.5\linewidth]{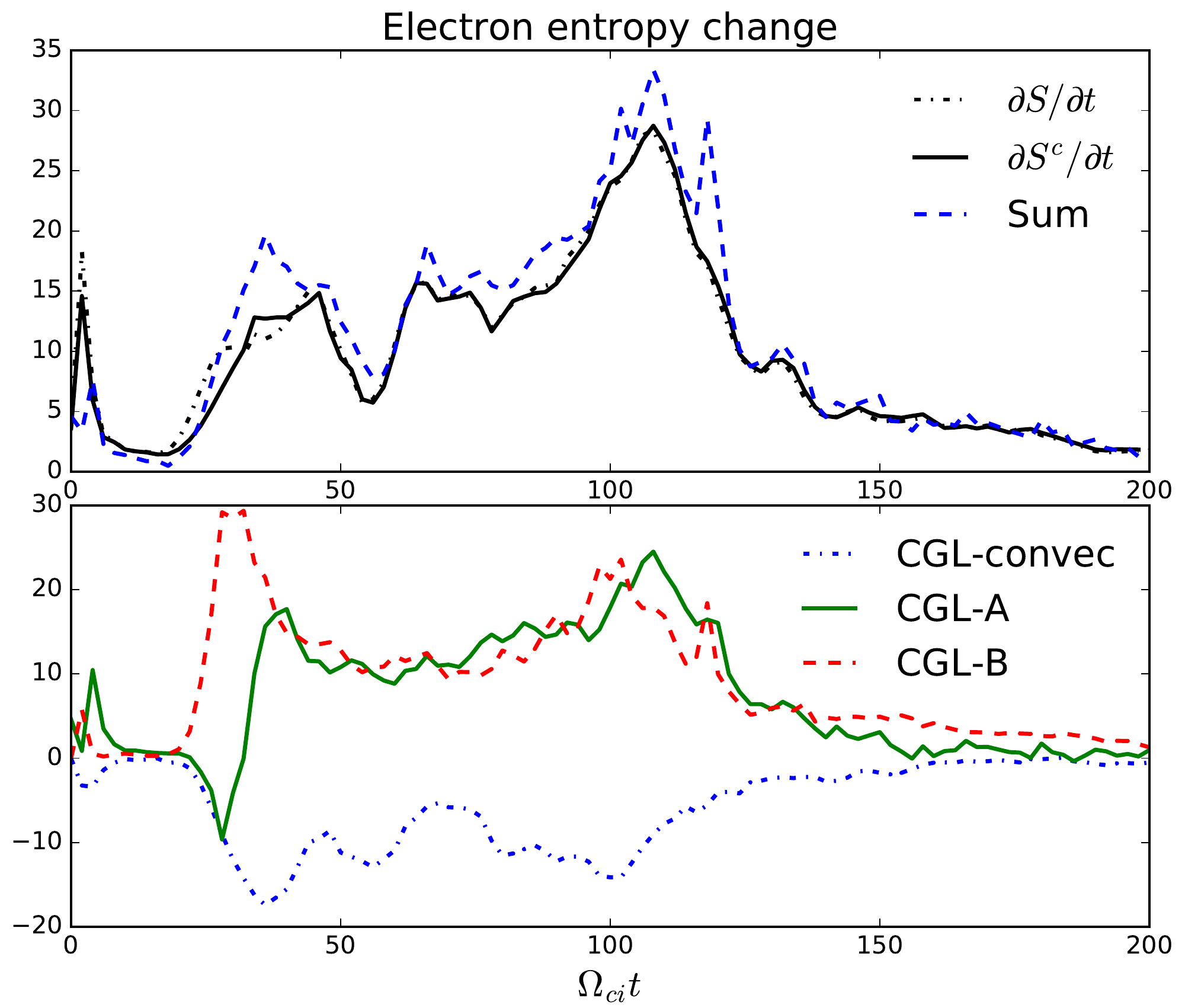}%
  \includegraphics[width=0.5\linewidth]{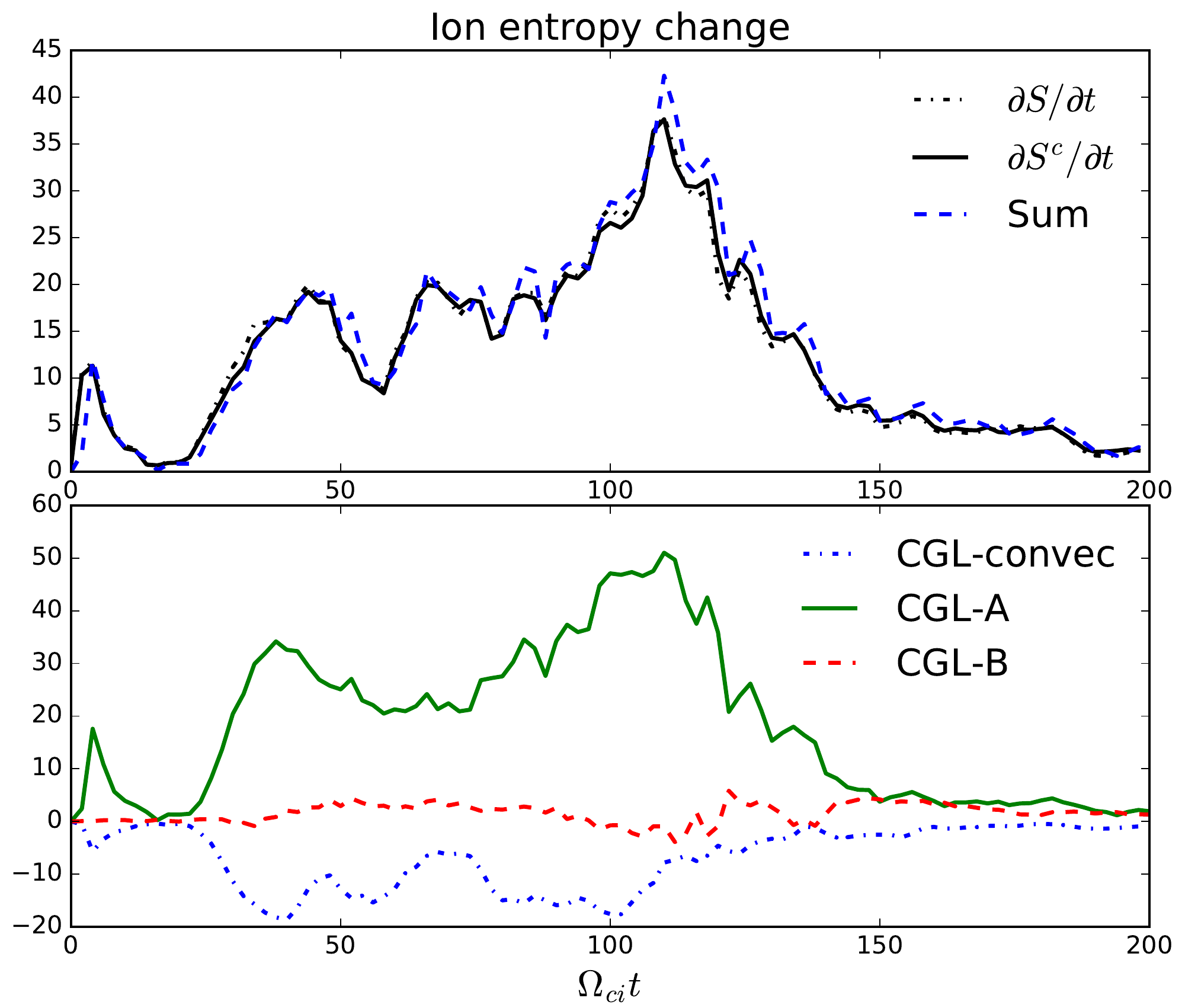}
  \caption{Rate of change of the CGL fluid entropy for electrons (left panels) and ions (right panels). The solid black lines in the top panels represent the rate of change of the CGL entropy $S^c$ while the dot-dashed black lines represent the regular fluid entropy $S$. The convection term and two other terms (see text for details) are evaluated separately and shown in the bottom panels. The sum of the three terms is displayed as the dashed blue curves in the top panels. \label{fig:dS-cgl}}
\end{figure*}

Similarly, we evaluate the CGL fluid entropy using Equation \eqref{eq:dSdt-CGL}, as shown in Figure \ref{fig:dS-cgl}. Similar to Figure \ref{fig:dS-iso}, the rate of change of the CGL entropy $S^c$ is displayed in the top panels as solid black lines. As a comparison, the rate of change of the isotropic fluid entropy $S$ is shown as dot-dashed lines (which is the same as the solid black curves in Figure \ref{fig:dS-iso}). The result shows that the two quantities $S$ and $S^c$ differ only slightly, probably because the anisotropy is not very strong \citep{Birn2006}. In the bottom panels of Figure \ref{fig:dS-cgl}, we separate Equation \eqref{eq:dSdt-CGL} into three parts. The first part is the convection, which is very close to the convection term in Figure \ref{fig:dS-iso}. The other two parts are denoted by $A$ and $B$, where:
\[ A = -\frac{\boldP^{n}:\boldsigma}{2\pper} - \frac{\nabla\cdot\boldq}{\pper}; \]
\[ B = - \kl{\frac{1}{\ppar} - \frac{1}{\pper}}\klm{\boldbb:(\boldP^{n}\cdot\nabla\boldu) - \frac{1}{2}\frac{d}{dt}(\boldbb):\boldP^{n} + \frac{1}{2}\boldbb:(\nabla\cdot\boldQ)}. \]
Term $A$ corresponds to the sum of Pi-D (with nongyrotropic pressure only) and heat flux, and $B$ is the additional term unique to the CGL entropy Equation \eqref{eq:dSdt-CGL}. Note that in calculating $d\boldb/dt$, we use the Faraday's law
\[ \partt{\boldB} = c\nabla\times\boldE \]
for convenience since this eliminates the need for evaluating time derivatives. The sum of the three parts is shown as dashed blue lines in the top panels, and they agree reasonably well with the solid black curves, though not as good compared to Figure \ref{fig:dS-iso}. This may be caused by the use of electric field and higher moments, which tend to be more noisy. One interesting difference between electrons and ions is that terms $A$ and $B$ are comparable in size for electrons while term $A$ dominates for ions. This may be understood from recognizing that Pi-D in Equation \eqref{eq:dSdt-iso} consists of both gyrotropic and nongyrotropic pressure while Pi-D in term $A$ consists of the nongyrotropic pressure only. Since electrons are strongly magnetized, the electron pressure is almost gyrotropic and thus term $A$ is dominated by the heat flux. As a result, for a weakly anisotropic pressure ($\ppar \simeq \pper$), the reduction of of Pi-D needs to be compensated by term $B$, which is dominated by the heat flux tensor $\boldQ$. This result demonstrates that even though the isotropic and CGL fluid entropy have similar values, their production mechanisms could be different.

\section{Discussion and conclusions}\label{sec:discussion}

In this paper, we discuss the evolution of the commonly used fluid entropy in both isotropic and gyrotropic (or CGL) forms. The isotropic fluid entropy is conserved for ideal gas or ideal MHD, and the CGL fluid entropy is conserved within the CGL plasma model. By simply taking moments of the Vlasov equation, we show that the fluid entropy of either form is not necessarily conserved even for an isolated collisionless system. This result is confirmed by a collisionless PIC simulation of multiple reconnecting current sheets. As pointed out by Liang et al.\cite{Liang2019}, kinetic entropy in a collisionless PIC simulation is approximately conserved. Clearly, the true Boltzmann entropy cannot be fully represented by a finite number of fluid moments. The increasing entropy in our simulation suggests that the fluid entropy is insufficient to capture the physical dissipation process. Instead, the change in fluid entropy may simply imply the breakdown of the ideal gas equation of state or the CGL double adiabatic equations.

The pressure-strain interaction or Pi-D has been proposed to capture the dissipation processes in space plasma \cite{Yang2017}. We illustrate in this paper that Pi-D does contribute to the change of fluid entropy in its isotropic form. However, our result shows that only the nongyrotropic part of Pi-D contributes to the CGL-form fluid entropy. Although the heat flux does not appear in the thermal energy equation, it plays an important role in the production of fluid entropy. Indeed, our simulation results suggest that the heat flux is almost equally as important as Pi-D in terms of increasing fluid entropy. When the CGL entropy is considered, the role of Pi-D is further reduced and the heat flux vector and tensor dominate the entropy increase, especially for electrons. Therefore, it may be expected that whether Pi-D or heat flux or other higher moments contribute to fluid entropy change depends on how the entropy itself is constructed.

Finally, we note that we do not completely rule out the possibility that the increase of fluid entropy is due to numerical issues (at least partly). A good way to clarify this will be to combine our analysis with the evaluation of kinetic entropy. This will be the goal of a future study.

\begin{acknowledgments}
SD and GPZ acknowledge the partial support of the NSF EPSCoR RII-Track-1 Cooperative Agreement OIA-1655280, and partial support from an NSF/DOE Partnership in Basic Plasma Science and Engineering via NSF grant PHY-1707247. XL acknowledge the support by NASA under grant NNH16AC60I, DOE OFES, and the support by the DOE through the LDRD program at Los Alamos National Laboratory (LANL). FG’s contributions are in part based upon work supported by the U.S. Department of Energy, Office of Fusion Energy Science, under Award Number DE-SC0018240 and by NSF grant AST-1735414. The simulation was performed at LANL Institutional Computing.
\end{acknowledgments}

\appendix*
\section{Parallel and perpendicular work}
Here, we show the relation between the parallel/perpendicular work \eqref{eq:work-parper} and the pressure equations \eqref{eq:ppar} and \eqref{eq:pper}. Using the ideal MHD induction equation, the parallel and perpendicular pressure equations become
\[ \frac{d\ppar}{dt} + \ppar\nabla\cdot\boldu + 2\ppar\kl{\frac{1}{B}\frac{dB}{dt} + \nabla\cdot\boldu} = 0; \]
\[ \frac{d\pper}{dt} + 2 \pper\nabla\cdot\boldu - \pper\kl{\frac{1}{B}\frac{dB}{dt} + \nabla\cdot\boldu} = 0. \]
Since $\ppar = 2 \varepsilon_{\parallel}$ and $\pper = \varepsilon_{\perp}$, we find the energy equations
\[ \partt{\varepsilon_{\parallel}} + \nabla\cdot(\varepsilon_{\parallel}\boldu) = - \ppar\frac{1}{B}\frac{dB}{dt} - \ppar \nabla\cdot\boldu; \]
\[ \partt{\varepsilon_{\perp}} + \nabla\cdot(\varepsilon_{\perp}\boldu) = \pper\frac{1}{B}\frac{dB}{dt}. \]
The term $\ppar\nabla\cdot\boldu$ on the RHS corresponds to the familiar $pdV$ work as
\[ \ppar\nabla\cdot\boldu = -\ppar\frac{1}{n}\frac{dn}{dt} = \ppar \frac{1}{V}\frac{dV}{dt} \quad\Rightarrow\quad \ppar dV = \ppar V(\nabla\cdot\boldu) dt. \]
Thus we multiply $Vdt$ on the RHS of the energy equations to obtain the work. The magnetic field related terms exchange parallel and perpendicular energy, and they provide additional work as
\[ dW_{B\parallel} = \ppar V \frac{dB}{B};\quad dW_{B\perp} = -\pper V \frac{dB}{B}. \]
Hence we establish the connection between Equations \eqref{eq:ppar}, \eqref{eq:pper} and \eqref{eq:work-parper}.

%% Create the reference section using BibTeX:

% \bibliography{bib2019}

\end{document}